\documentclass[conference]{IEEEtran}
\IEEEoverridecommandlockouts

\usepackage{cite}

\ifCLASSINFOpdf
  \usepackage[pdftex]{graphicx}
\else
\fi

\usepackage{amssymb}
\usepackage{xcolor}
\usepackage{amsmath}
\usepackage{epsfig}
\usepackage{comment}
\usepackage{subfigure}
\usepackage{subcaption}
\usepackage{multirow}
\usepackage{makecell}
\usepackage{array}
\usepackage[lofdepth,lotdepth]{subfig}
\usepackage{subcaption}
\usepackage{graphicx}

\begin{document}

\title{RIS Size Determination Across Frequencies and Deployment Scenarios: A Simulation-Based Study}





\author{\IEEEauthorblockN{Emre ARSLAN\IEEEauthorrefmark{1}\IEEEauthorrefmark{2}, Ahmet Faruk COŞKUN\IEEEauthorrefmark{1}}
\IEEEauthorblockA{\IEEEauthorrefmark{1}6GEN Laboratory, Next-Generation R\&D, Network Technologies, Turkcell, {\.{I}}stanbul, Turkiye}

\IEEEauthorblockA{\IEEEauthorrefmark{2}Department of Electrical and Electronics Engineering, Koç University, {\.{I}}stanbul, Turkiye\\ Emails: {\{emre.arslan, coskun.ahmet\}@turkcell.com.tr}}}
\maketitle

\begin{abstract}
Despite the growing interest in the integration of reconfigurable intelligent surfaces (RIS) into next-generation wireless communications systems, a critical gap remains in understanding what the dimensions of an RIS must be to provide meaningful performance gains across realistic deployment scenarios. This paper addresses this challenge by presenting a practical and scenario-aware methodology for determining optimal RIS dimensions, tailored to specific frequency bands, environments, and use cases. Leveraging a realistic simulation model that incorporates angular scattering characteristics, practical network node locations, and propagation constraints, we evaluate the RIS-assisted performance in a diverse set of configurations. For selected use-cases, we quantify key performance indicators such as average signal-to-noise ratio and outage probability, and we demonstrate how RIS size impacts system reliability. Our findings show that RIS deployment effectiveness is highly sensitive to both physical size and geometric placement, and that there is no one-size-fits-all solution. The proposed framework, supported by detailed use case tables and validated through comprehensive simulations, offers design guidelines for operators and vendors seeking to deploy RIS in practical wireless network settings.

\end{abstract}


\section{Introduction}
Reconfigurable intelligent surfaces (RIS) have emerged as one of the most promising enabling technologies of future wireless communication systems \cite{6G}. While the fundamental principles and theoretical gains of RIS have been well-studied \cite{RIS, Ellingson2021, Basar2021, Tang2021}, a critical and often underexplored challenge lies in practical RIS deployments. One of the most decisive factors influencing RIS performance is its physical size that is, the number of reflective elements and the total surface area. However, there is no common RIS size that can guarantee satisfactory performance across all deployment scenarios. The optimal RIS dimensions are highly dependent on the specific use case, which may vary significantly in terms of environment (indoor vs. outdoor), frequency band (sub-6 GHz, FR3, FR2), base station (BS)-RIS-user equipment (UE) topology. RIS use cases range from short-range indoor mmWave deployments to large-scale rural macrocell coverage extensions in FR3 bands. Each use case exhibits distinct propagation characteristics, angular spreads, path loss conditions, and practical constraints. Consequently, the required RIS size for achieving a target signal-to-noise ratio (SNR), throughput, or outage probability (OP) can differ drastically. Deploying an excessively large RIS in a scenario where modest gains are already achievable would result in unnecessary costs and complexity, especially considering the production and operational costs 
associated with large RIS panels. Conversely, a very small-sized RIS that is too small might fail to deliver the expected performance benefits for a challenging use case
, thereby negating the very purpose of using RIS in the first place.

Despite the clear importance of aligning RIS design with the deployment scenario, the existing literature is limited in its ability to provide realistic, scenario-specific sizing guidelines. Many studies rely on overly idealized channel models, assume perfect coverage conditions, or do not account for angular sensitivity and geometric placement impacts. Furthermore, few works systematically evaluate how RIS dimensions influence system performance across multiple use cases using practical scattering models. As RIS prototypes move closer to real-world deployment and standardization discussions intensify, there is an urgent need for tools and studies that bridge the gap between theoretical gains and deployment realities.

In this study, we address this gap by proposing a practical and use-case-aware methodology for determining the required RIS size under varying deployment conditions. Our approach incorporates realistic simulation models that account for angular scattering, and scenario-specific node locations. We analyze how different RIS sizes affect key performance indicators such as received power, SNR, and OP. By systematically applying this framework across a diverse set of use cases including indoor mmWave, urban Sub-6 GHz microcells, and rural FR3 macrocells, we demonstrate that RIS dimensioning is far from a one-size-fits-all problem. Through detailed simulations and statistical analysis, we identify the minimum RIS sizes needed to meet target SNR thresholds and ensure acceptable outage probabilities. The resulting insights are summarized in performance figures and tables, which offer practical sizing guidelines for network planners and infrastructure designers. 

Section II presents the methodology on how to determine the RIS dimensions for the usecases, the signal model utilized for the RIS-assisted communications, and its validation. Numerical results based on the simulation set-up are presented in Section III. Finally, Section IV concludes the paper and discusses future open areas for investigation.

\section{Determining RIS Dimensions}

\subsection{Methodology}
\begin{figure*}
    \centering
    \subfigure[Simulation scenario: 3D view]
    {\includegraphics[width=0.78\columnwidth]{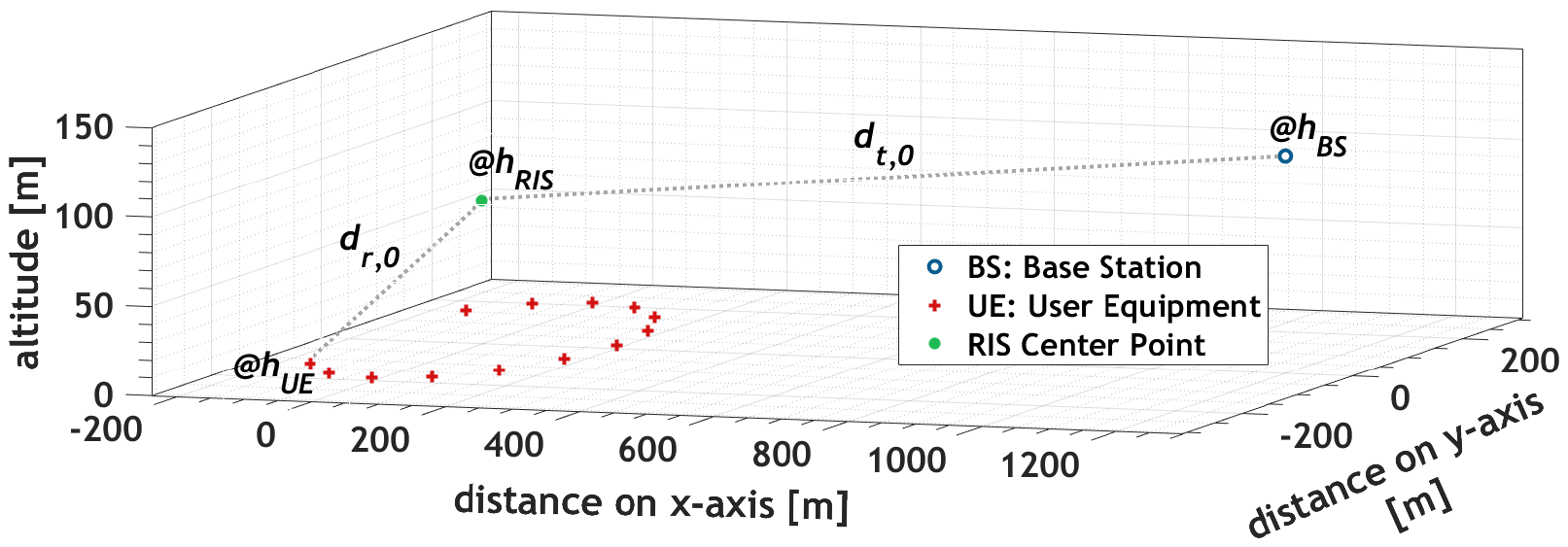}}
    \subfigure[Simulation scenario: 2D top-view]{ \includegraphics[width=0.65\columnwidth]{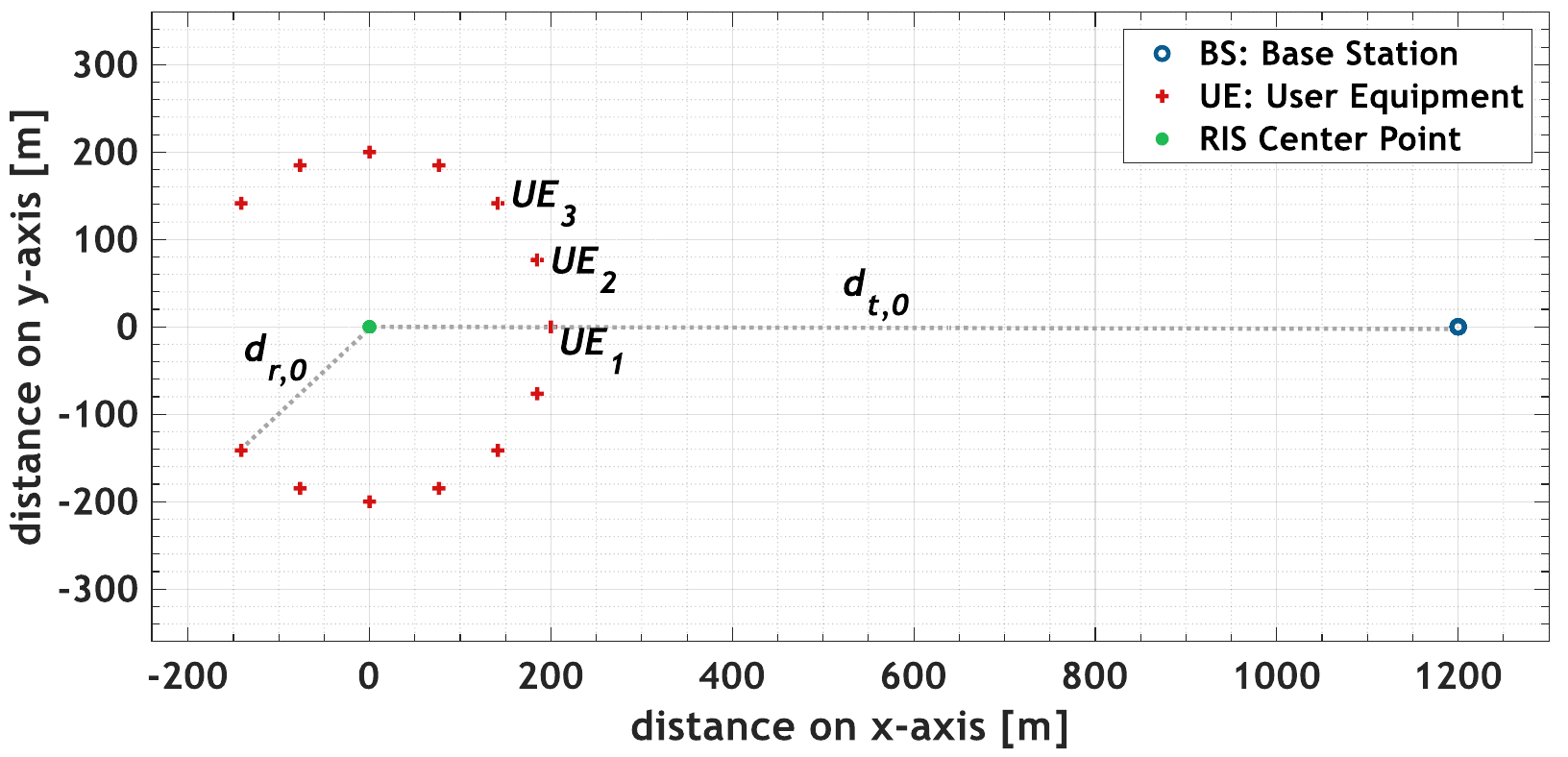}}
    \subfigure[Average SNR for UE bearings]{ \includegraphics[width=0.55\columnwidth]{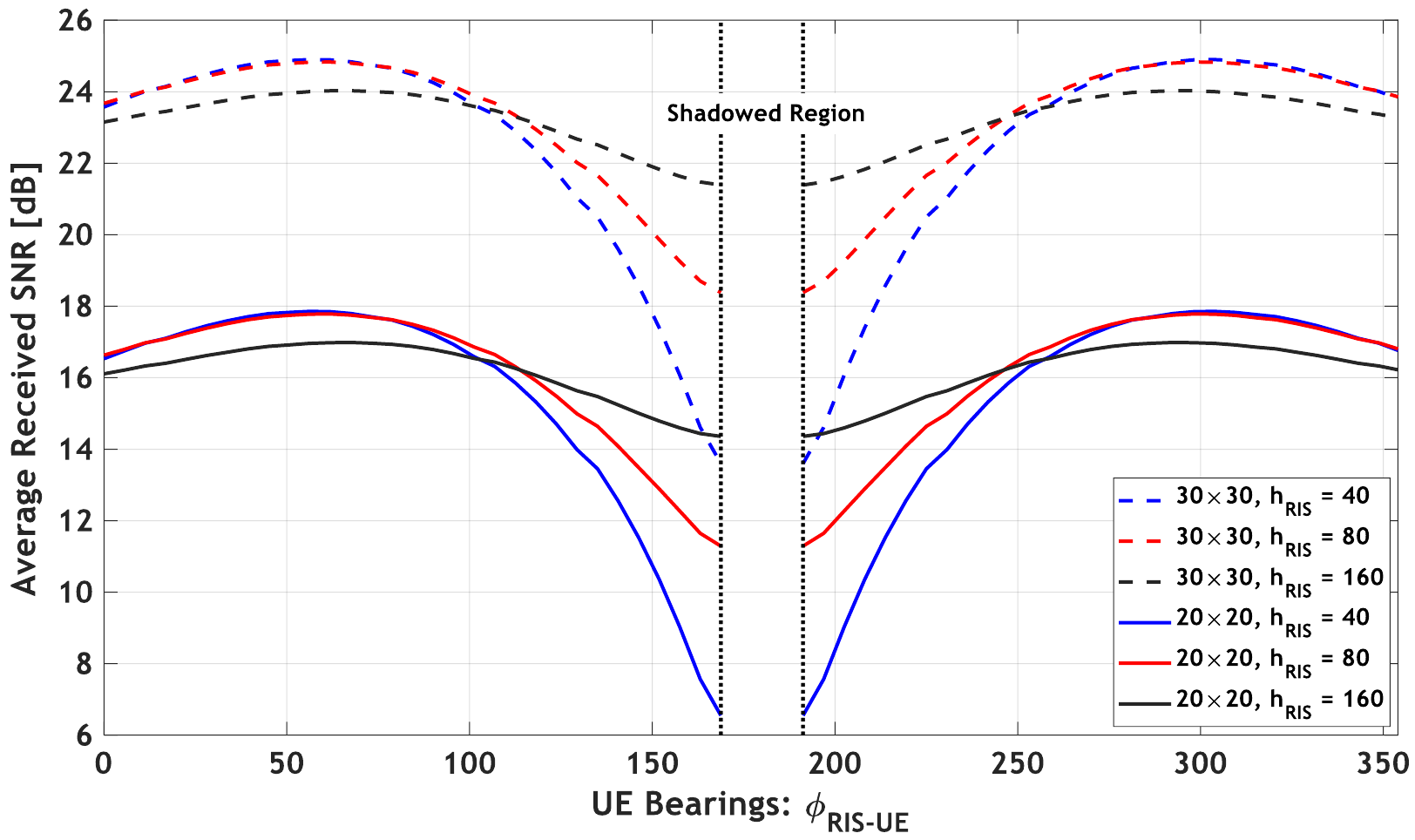}}
    
    \caption{Simulation scenario and locations of the network nodes}
    \label{fig:SimulationScene}
\end{figure*}

This subsection introduces the systematic methodology used to determine the optimal physical dimensions of RIS across a wide range of frequencies and deployment scenarios. The methodology is based on a realistic simulation framework designed to account for practical scattering behavior and environmental variability. As a foundation, we define a generic RIS-assisted communication scenario, illustrated in Fig. \ref{fig:SimulationScene}-(a) and (b), which models a transmitter, a receiver, and a strategically-placed RIS to enhance signal quality via scattered paths. This template scenario serves as a baseline that we adapt to various use cases, including indoor, outdoor, urban, rural, macrocell, and microcell configurations, as well as different frequency bands such as sub-6 GHz, mmWave, WiFi, and FR3.
By realizing the template scenario with the simulation parameters given in Table \ref{SimulationParameters}, and the parameters given in \cite[Table I]{AFC_EA}, the variation of average RIS-assisted SNR has been depicted for different RIS sizes and different RIS altitudes in Fig. \ref{fig:SimulationScene}-(c). Here, the curves clearly show the signal focusing capability of RIS w.r.t. relative azimuth target directions, and also exhibit the dependency of the overall SNR performance to the variation in RIS altitude.

Fig. \ref{fig:SimulationScene}-(c) presents the average received power at possible UE locations. It can be seen that when the UE is not optimally placed to obtain maximum received power, the received power levels may fluctuate within a $7$ dB margin which is a significant reduction. Hence, we can conclude that the relative azimuth angle of the UE plays a significant role in overall performance metrics. It is not enough to only assume that the RIS could direct the impinging signal to wherever it likes via phase shifts. One must also consider where the signals impinging are coming from and where the UE is located for the RIS to perform redirection for maximal performance gains. The common research studies with simulation, mathematical analysis and field trial perspectives do not take the varying scattering capability of reconfigurable surfaces into account, which might actually cause overestimating the advantages of RIS-assisted communications scenarios. Fig. \ref{fig:SimulationScene}-(c) shows that while increasing the RIS size leads to a notable improvement in the average received SNR, the overall performance trend across different UE placements remains relatively consistent. A key observation is that the RIS deployment height significantly affects how sensitive the performance is to UE positioning. At a lower RIS height of $40$m, there is considerable variation in SNR across different UE bearings, whereas at $160$m, this variation diminishes, indicating more uniform coverage and reduced dependence on exact UE location. This difference arises because lower RIS placements restrict the angular visibility and beamforming coverage toward UEs at wider bearings, making the system more susceptible to geometric misalignment and angular mismatch, which are mitigated at higher RIS elevations.

\begin{table}[h]
  \centering
  \caption{\textsc{Simulation Parameters}}
  \label{SimulationParameters}
  \begin{tabular}{c c c}
    \hline
    {\bf{Param.}} & {\bf{Definition}} & {\bf{Value}} \\
    \hline\hline
     $f_c$ & Center frequency & $3500$ MHz \\\hline
    $BW$ & Signal bandwidth & $15$ kHz \\\hline
    $P_t$ & Transmit Power & $20$ Watts \\\hline
    $N_h$ & \#RIS unit cells in horizontal axis & $\{20, 30\}$ \\\hline
    $N_v$ & \#RIS unit cells in vertical axis & $\{20, 30\}$ \\\hline
    $\Delta h$ & height of RIS unit cell & $3\lambda/7$ [m] \\\hline
    $\Delta v$ & width of RIS unit cell & $3\lambda/7$ [m] \\\hline
    $d_h$ & hor. spacing between unit cells & $\lambda/2$ [m] \\\hline
    $d_v$ & ver. spacing between unit cells & $\lambda/2$ [m] \\\hline
    $K_{\Phi}$ & \#RIS phase states& $2$\\\hline
    BS$\left(x, y, z\right)$ & BS Location & $\left(1200, 0, 122\right)$ \\\hline
    RIS$\left(x, y, z\right)$ & RIS Center Point & $\left(0, 0, 80\right)$ \\\hline
    $h_{\text{UE}}$ & altitude of UE terminals & $2$ m \\\hline
    $d_{t,0}$ & distance between BS and RIS & $1200$ m \\\hline
    $d_{r,0}$ & distance between RIS and EUs & $200$ m \\\hline
  \end{tabular}
\end{table}

For each use case, we generate a large pool of simulation data that captures all possible RIS-assisted signal power realizations encountered from different communications scenarios. 
Using this data pool, we extract the probability density functions (PDFs) 
of the scattering power and SNR for various RIS sizes. These distributions allow us to quantify the number of reflective elements over which the RIS will 
satisfy a predefined SNR limit (e.g., $30$ dB), and/or maintain the communications below an OP boundary. This approach provides a concrete, data-driven basis for RIS design decisions and helps guide real-world deployments across several RIS use cases.


\subsection{Signal Model For RIS aided-communication}

Throughout the simulation-based evaluations targeting the exhibition of RIS size requirements, the RIS-assisted signal model introduced in \cite{AFC_EA} has been utilized. By assuming a perfect obstruction (i.e., non line-of-sight, NLoS) between BS and UE points, the received baseband signal at the discrete time instant $n$ can be expressed as follows:
\begin{flalign}
r[n] =\sum_{k=1}^{N_v}\sum_{l=1}^{N_h}\rho_{kl}{e^{-j\Psi_{kl}}}s[n-n_{kl}] + w[n].
\end{flalign}
where the parameters in eq. (1) are available in \cite[Section II and III]{AFC_EA}. After having defined the instantaneous bi-static scattered signal for a given RIS-assisted wireless communications scenario, the corresponding instantaneous SNR might be evaluated by using the proportion of the instantaneous received signal power $\mu[n]\triangleq \left[|y[n]|^2\right]$ and the noise variance at the UE antenna $(N_o\triangleq k_BT_sBWF_n)$: 
$\gamma[n]\triangleq \mu[n]{E_b}/{N_o}$.
Here, $y[n]$ is the signal part of $r[n]$, $k_B$ is the Boltzmann constant, $T_s$ is the system temperature in Kelvin, $BW$ is the UE bandwidth, and $F_n$ is the noise figure. 

\subsection{Validation}
\begin{figure}
    \subfigure[Schematic view of nodes]
    {\includegraphics[width=0.47\columnwidth]{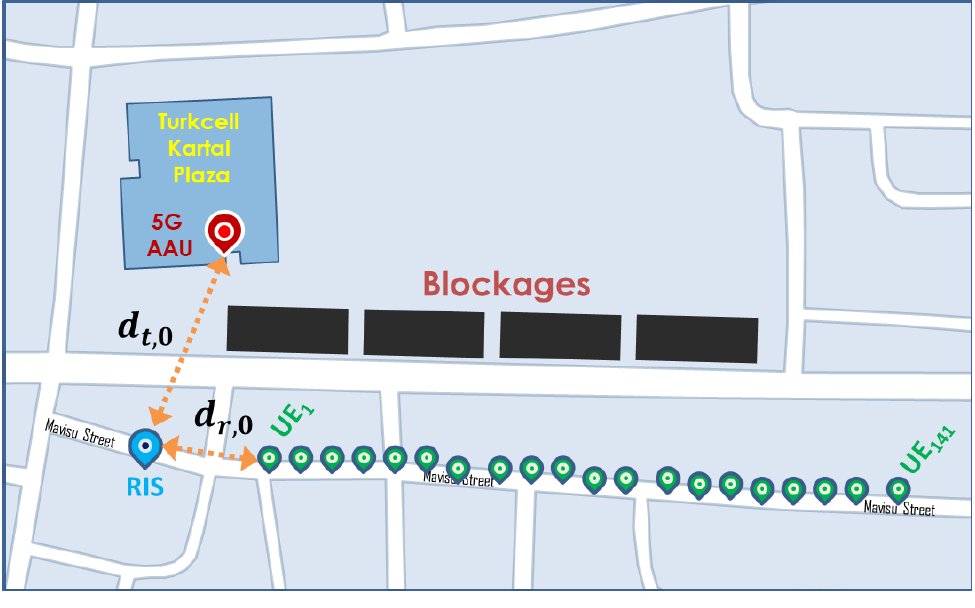}}
    \subfigure[Average SNR comparison]{ \includegraphics[width=0.51\columnwidth]{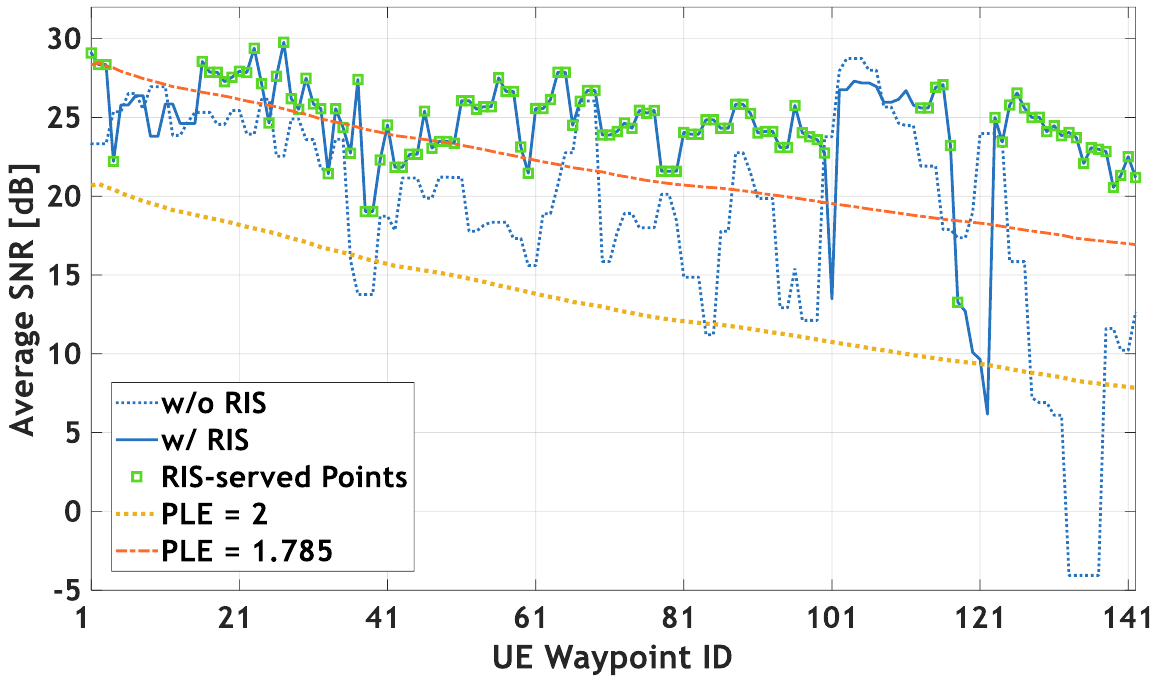}}
    \caption{Validation scenario and results}
    \label{fig:validationScenario}
\end{figure}

This subsection provides the outcomes of a brief validation study that will be critical to rely on the following outcomes of the basic RIS size determination study. In order to validate the proposed simulator's evaluations, the measurement records for a fixed BS, RIS and UE trajectory which is depicted in Fig. \ref{fig:validationScenario}-(a) have been utilized. The experimental study has been performed in Turkcell Kartal Plaza by using the mmWave RIS product of ZTE Corporation \cite{RISPR}.

In the measurement study, the BS was mounted on a tower at an altitude of $88.7$ meters and operated at a center frequency of $26$ GHz. The RIS product of ZTE (i.e., Dynamic RIS 2.0) was installed $80$ meters distant from the BS on a second-floor balcony at an altitude of $64$ meters. The UE test locations sampled with totally $141$ coordinate points (corresponding to distances up to $205$ meters w.r.t. RIS center) were carefully chosen to ensure LoS with the RIS while being in NLoS relative to the BS. 
The experiments demonstrated that the RIS effectively extended coverage, achieving a maximum bi-static range of $280$ meters (BS-RIS-UE). As seen from the average SNR curve given in Fig. \ref{fig:validationScenario}-(b) in green, the RIS-assistance provides considerable gains when compared to the ``w/o RIS" case. By substituting the UE coordinates recorded during the test into the simulation variables together with the BS and RIS coordinates and the technical parameters (e.g., number of RIS elements, element spacing, etc.), the corresponding average SNR values for each UE location have been evaluated under the free-space propagation assumption. To provide a theoretical/simulation-based benchmark, the simulation framework has focused on the generic path-loss exponent (PLE) of $2$, and a more practical PLE of $1.785$ which has been empirically shown in \cite{PLERef} to be more suitable for mmWave propagation. 
The average SNR curves given in yellow and orange represent the main behavior for $\emph{\emph{PLE}}=2$ and $\emph{\emph{PLE}}=1.785$, where the latter is seen to better follow the main characteristics of the measurement record.
Despite the variations between the orange curve (i.e., theoretical) and the green curve (i.e., measurement record) due to the channel fluctuations that cannot be accessed during field tests, the consistency in the main trend confirms the simulator’s ability to accurately model RIS-assisted propagation scenarios.

\subsection{Set Up Explanations}
The simulation setup in this study is constructed based on the generic RIS-assisted communication template illustrated in Fig. \ref{fig:SimulationScene}-(a) and (b). This template defines the spatial configuration of the BS, UE, and RIS, along with key geometric parameters such as distances, angles of incidence and reflection, and RIS placement. The setup is designed to be flexible and adaptable, allowing us to simulate a wide range of deployment scenarios by varying the propagation environment, frequency band, and physical layout. It should be noted that the shadowed region exists since the considered RIS is not transmissive and cannot reflect signals thorough/behind itself. More comprehensive details regarding the simulation set-up can be found in \cite{AFC_EA}.

The combination of the generalized template and the structured use case definitions ensures consistency across simulations while allowing meaningful comparisons between scenarios. By performing successive simulation runs that take all the parameters of each use-case (given in Tables \ref{table:UCParameters1}, \ref{table:UCParameters2}, \ref{table:UCParameters3}) as inputs, average values for RIS-assisted received power and SNR have been evaluated, and exhibited for a given RIS size.

\renewcommand{\arraystretch}{1.5}
\begin{table*}[t]
\caption{RIS Use-Case Outdoor Parameters}
\centering
\label{table:UCParameters1}
\begin{tabular}{c | c c c c c c c c c}
\hline
\hline
Use-Case ID & \thead{1} & \thead{2} & \thead{3} & \thead{4} & \thead{5}\\
\thead{Use-Case \\Definition} & \thead{Sub-6 \\ Umi} & \thead{Sub-6 \\ Uma} & \thead{Sub-6 \\ RMa} & \thead{FR3 \\ Umi} & \thead{FR3 \\ Uma}             \\ \hline\hline
$P_{t,\text{BS}}$ (dBm) & 37 & 46 & 52 & 37 & 46   \\ \hline
$f_{c,\text{BS}}$ (MHz) & 3500 & 3500 & 3500 & 8000 & 8000  \\ \hline
$B_{\text{UE}}$ (kHz) & 1440 & 1440 & 1440 & 2880 & 2880   \\ \hline
$\mathbf{h}_{\text{BS}} [m]$ & \{10 15 20 30\} & \{10 15 20 30\} & \{10 15 20 30\} & \{10 15 20 30\} & \{10 15 20 30\}  \\ \hline
$\mathbf{h}_{\text{RIS}} [m]$ & \{10 25 40\} & \{10 25 40\} & \{10 25 40\} & \{10 25 40\} & \{10 25 40\}   \\ \hline
$\mathbf{h}_{\text{UE}} [m]$ & \{1 2 5 15 30 50 100\} & \{1 2 5 15 30 50 100\}
& \{1 2 5 15 30\} & \{1 2 5 15 30 50 100\} & \{1 2 5 15 30 50 100\}  
\\ \hline
$\mathbf{d}_{\text{BS-RIS}} [m]$ & \{50 100 150 200\}                    & \{200 300 400 500\} & \{500 1000 1500 2000\} & \{50 75 125 150\} & \{100 175 225 300\}    \\ \hline
$\mathbf{d}_{\text{RIS-UE}} [m]$ & \{50 100 150 200\}                    & \{200 300 400 500\} & \{500 1000 1500 2000\} & \{50 75 125 150\} & \{50 75 125 150\}   \\ \hline\hline
\end{tabular}
\end{table*}

\renewcommand{\arraystretch}{1.5}
\begin{table*}[t]
\caption{RIS Use-Case Outdoor Parameters (Continued)}
\centering
\label{table:UCParameters2}
\begin{tabular}{c | c c c c c c c c c}
\hline
\hline
Use-Case ID & \thead{6} & \thead{7} & \thead{8} & \thead{9}\\
\thead{Use-Case\\Definition} & \thead{FR3 \\ RMa} & \thead{mmW \\ Umi} & \thead{mmW \\ Uma} & \thead{mmW \\ RMa}               \\ \hline\hline
$P_{t,\text{BS}}$ (dBm) & 52 & 40 & 49 & 55  \\ \hline
$f_{c,\text{BS}}$ (MHz) & 8000 & 27000 & 27000 & 27000 \\ \hline
$B_{\text{UE}}$ (kHz) & 2880 & 8640 & 8640 & 17280 \\ \hline
$\mathbf{h}_{\text{BS}} [m]$ & \{10 15 20 30\} & \{10 15 20 30\} & \{10 15 20 30\} & \{10 15 20 30\} \\ \hline
$\mathbf{h}_{\text{RIS}} [m]$ & \{10 25 40\} & \{10 25 40\} &  \{10 25 40\} &  \{10 25 40\}  \\ \hline
$\mathbf{h}_{\text{UE}} [m]$ & \{1 2 5 15 30\} & \{1 2 5 15 30 50 100\} & \{1 2 5 15 30 50 100\} & \{1 2 5 15 30\}
\\ \hline
$\mathbf{d}_{\text{BS-RIS}} [m]$ & \{300 700 1100 1500\} &\{20 50 70 100\} & \{50 100 150 200\} & \{100 400 700 1000\}     \\ \hline
$\mathbf{d}_{\text{RIS-UE}} [m]$ & \{300 700 1100 1500\}   & \{20 50 70 100\} & \{50 100 150 200\} & \{100 400 700 1000\}         \\ \hline\hline
\end{tabular}
\end{table*}

\renewcommand{\arraystretch}{1.5}
\begin{table*}[t]
\caption{RIS Use-Case Indoor Parameters}
\centering
\label{table:UCParameters3}
\begin{tabular}{c | c c c c c c c c c}
\hline\hline
Use-Case ID & \thead{10} & \thead{11} & \thead{12} & \thead{13} & \thead{14} & \thead{15} & \thead{16}\\
\thead{Use-Case\\Definition} & Home Wifi                    & \thead{Sub-6 GHz \\ Small Office} & \thead{Sub-6 GHz \\ Large Industrial} & \thead{FR3 \\ Small Office} & \thead{FR3 \\ Large Industrial} & \thead{mmWave \\ Small Office} & \thead{mmWave \\ Large Industrial}                 \\ \hline\hline 
$\mathbf{P}_{t,\text{BS}}$ (dBm) & 30 & 30 & 37 & 30 & 37 & 30 & 37
\\ \hline
$f_{c,\text{BS}}$ (MHz) & 6000 & 3500 & 3500 & 8000 & 8000 & 27000 & 27000
\\ \hline
$B_{\text{UE}}$ (kHz) & 2160 & 1440 & 2880 & 4320 & 8640 & 8640 & 8640
\\ \hline 
$\mathbf{h}_{\text{BS}} [m]$ & \{1 2 3\} & \{2 3 4\} & \{4 6 8 10\} & \{2 3 4\} & \{4 6 8 10\} & \{2 3 4\} & \{4 6 8 10\}
\\ \hline
$\mathbf{h}_{\text{RIS}} [m]$ & \{2 3\} & \{2 3 4\} & \{4 6 8 10\} & \{2 4 6\} & \{4 6 8 10\} & \{2 4 6\} & \{4 6 8 10\}  \\ \hline
$\mathbf{h}_{\text{UE}} [m]$ & \{1 2\} & \{1 2 3\} & \{1 2 4 6 8\} & \{1 2\} & \{1 2 4 6 8\} & \{1 2\} & \{1 2 4 6 8\}  \\ \hline
$\mathbf{d}_{\text{BS-RIS}} [m]$ & \{3 5 8\} & \{3 10 15 20\} & \{20 40 60 80\} & \{2 5 10 15\} & \{10 30 50 80\} & \{1 3 6.5 10\}    & \{5 20 50 80\}         \\ \hline
$\mathbf{d}_{\text{RIS-UE}} [m]$ & \{3 5 8\} & \{3 10 15 20\} & \{20 40 60 80\} & \{2 5 10 15\} & \{10 30 50 80\} & \{1 3 6.5 10\} & \{5 20 50 80\}           \\ \hline\hline
\end{tabular}
\end{table*}

\subsection{RIS Use Cases}

To effectively determine the practical RIS size requirements across diverse deployment environments, we extend our analysis beyond the generalized simulation template to a comprehensive set of scenario-specific use cases. As noted earlier, RIS-assisted communication performance is influenced by a broad range of parameters including operating frequency, BS-RIS and RIS-UE distances, heights of network elements, angular distributions, and deployment type (e.g., indoor, outdoor, urban, rural). To ensure realism, we define these parameters based on insights from existing literature, industry standards, and operator experience. 

Specifically, the use case parameters have been informed by well-known propagation studies and RIS channel modeling efforts such as \cite{Ellingson2021, Basar2021, Tang2021}, as well as mmWave and sub-6 GHz deployment guidelines presented in 3GPP TR 38.901 \cite{3gpp38901}, and TR 38.873 \cite{3gpp38873}. Parameters such as center frequency, transmit power, bandwidth, and antenna heights have been selected in line with typical values adopted in recent RIS research and system evaluations. In addition, we incorporate scenario-specific variations such as BS height distributions, RIS mounting positions, and UE altitudes that reflect practical constraints and operator-side deployment experience, drawing also on works such as \cite{3gpp38901, 3gpp38873}. The complete list of use cases, along with all key configuration parameters and their associated probability distributions, is presented in Table \ref{table:UCParameters1} (continued in Tables \ref{table:UCParameters2} and \ref{table:UCParameters3} for extended scenarios). These tables collectively span a wide spectrum of realistic environments, including outdoor urban micro (Umi), urban macro (Uma), and rural macro (RMa) setups across Sub-6 GHz, FR3, and mmWave bands, as well as indoor environments such as small offices, large industrial facilities, and home WiFi. Each table defines carrier frequency, BS and RIS heights, UE position ranges, BS-RIS and RIS-UE distances.

As illustrated in our preliminary results given in Fig. \ref{fig:validationScenario}-(c), each deployment scenario exhibits distinct propagation behavior, which in turn affects how much RIS aperture area is needed to satisfy target KPIs. Therefore, it is essential to perform a dedicated RIS size evaluation for each use case, as a one-size-fits-all approach would fail to capture the performance trade-offs unique to each frequency band and environment.

For each RIS size, we computed key performance indicators (KPIs) that reflect an aggregated outcome across multiple environmental and operational setups. Our proposed simulation setup thus serves as a comprehensive tool to estimate optimal RIS dimensions based on literature-supported use cases. Moreover, as a telecommunications operator with extensive experience in next-generation wireless network infrastructure, our parameter selections and scenario specifications are also grounded in practical network planning perspectives and real-world deployment expertise.

For detailed analysis, we select a representative subset of three use cases namely UC-5, UC-7, 
and UC-16, which exemplify the diversity of deployment conditions. For each use case, we apply the methodology introduced in Section II and compute average metrics based on the probability distributions of scenario parameters. This yields realistic performance estimates for RIS-assisted communications in the form of average scattered power, SNR, and OP.

\section{Numerical Results}
\begin{figure}
    \centering
    
    \subfigure[Received power for UC-5] {\includegraphics[width=0.48\columnwidth]{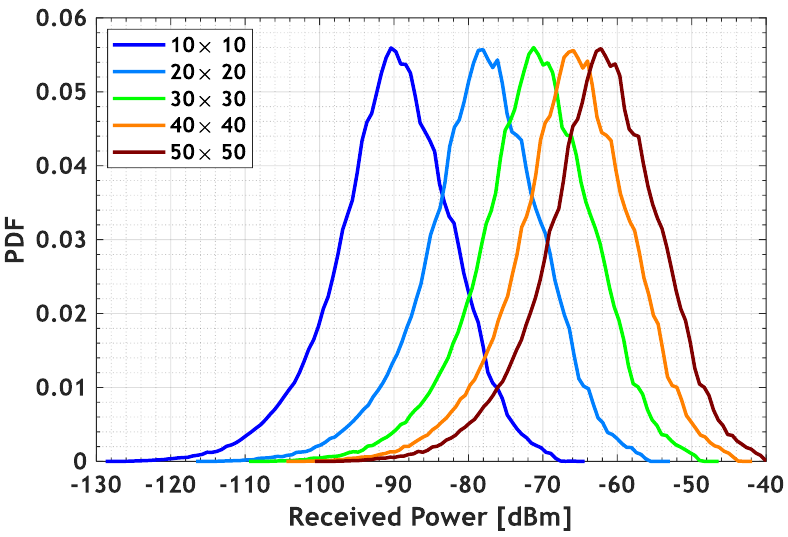}}
    \subfigure[Received SNR for UC-5]{ \includegraphics[width=0.48\columnwidth]{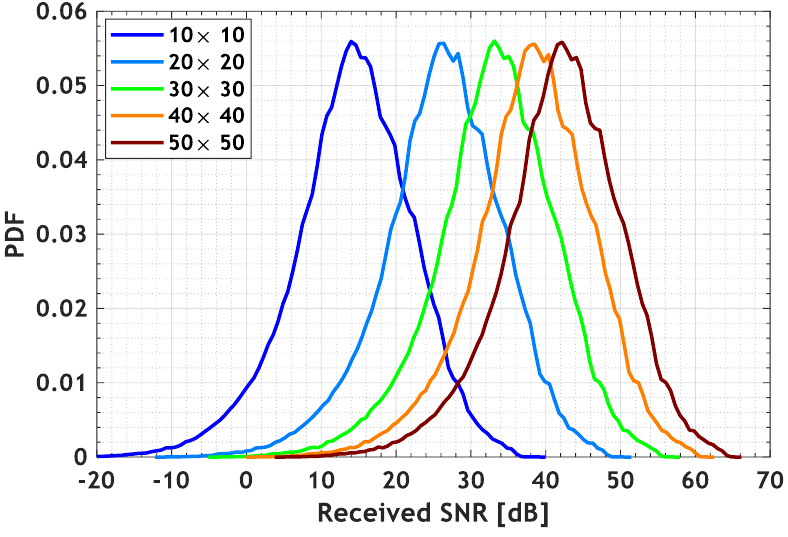}}
   
    \subfigure[Received power for UC-7]{\includegraphics[width=0.48\columnwidth]{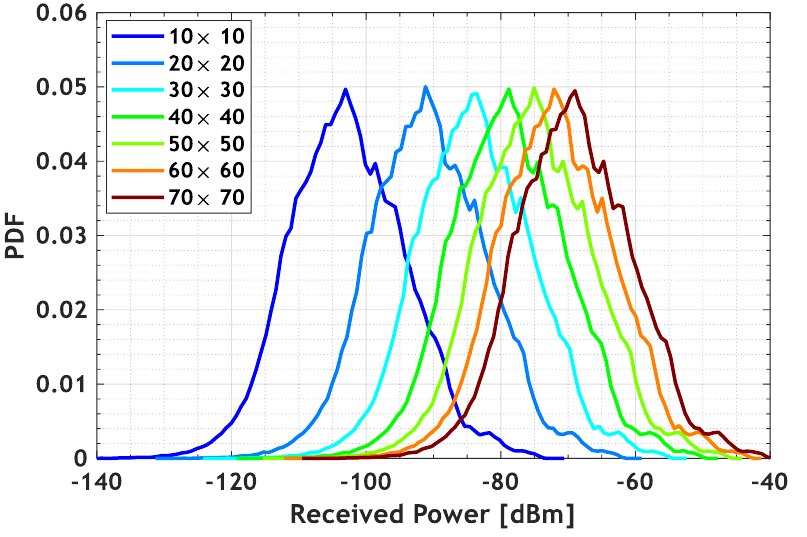}}
    \subfigure[Received SNR for UC-7]{ \includegraphics[width=0.48\columnwidth]{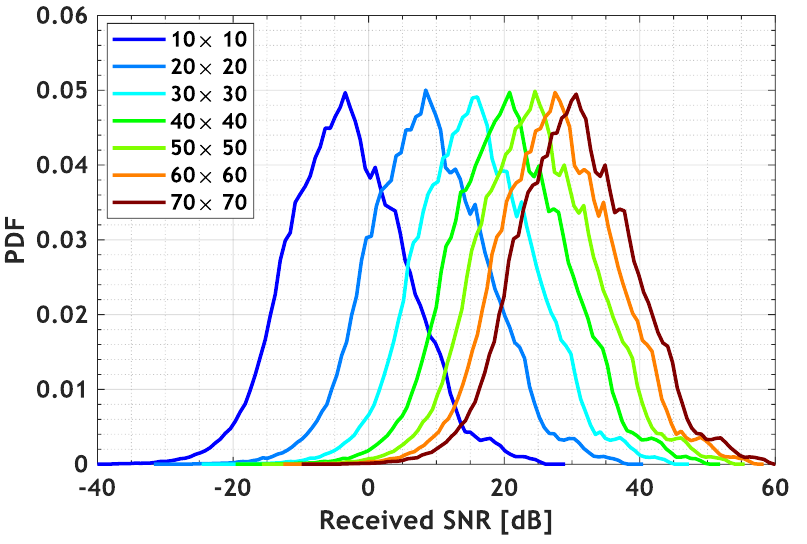}}
    

    \subfigure[Received power for UC-16]{\includegraphics[width=0.48\columnwidth]{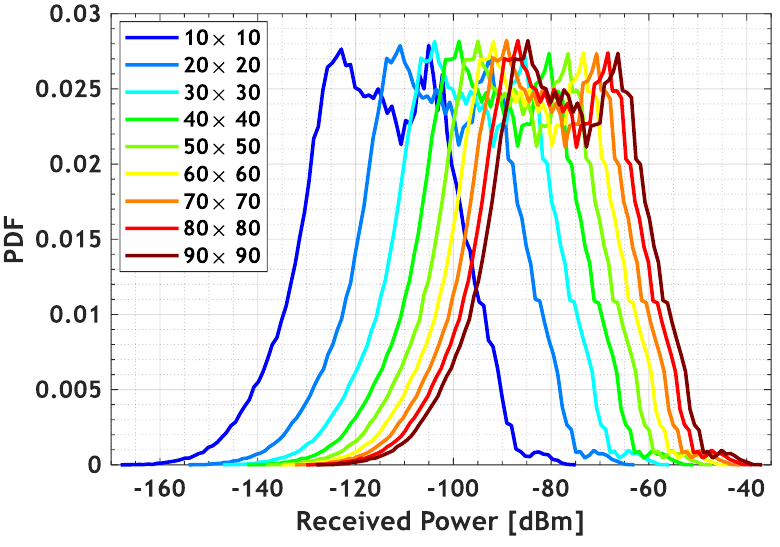}}
    \subfigure[Received SNR for UC-16]{ \includegraphics[width=0.48\columnwidth]{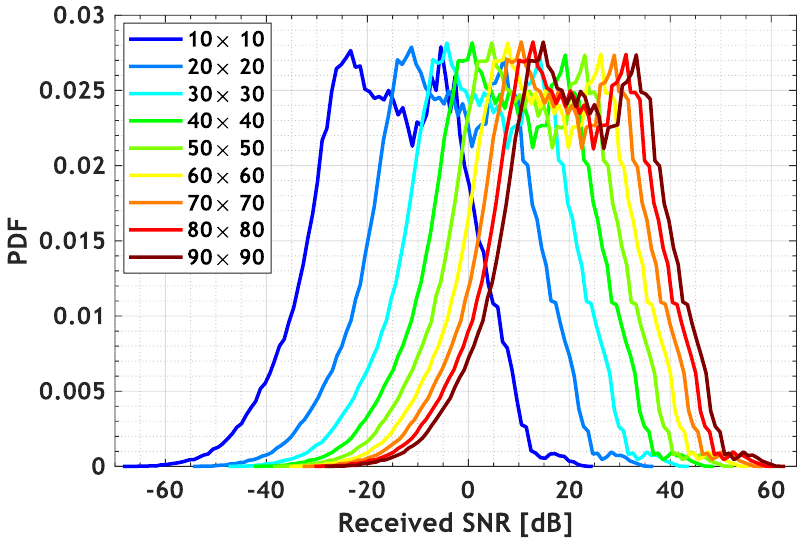}}
\caption{Probability densities of RIS-assisted signal power and SNR for selected UCs and various RIS sizes}
\label{fig:ProbabilityDensities}
\end{figure}

\begin{figure}[t]
    \centering
    \subfigure[UC-5] {\includegraphics[width=0.98\columnwidth]{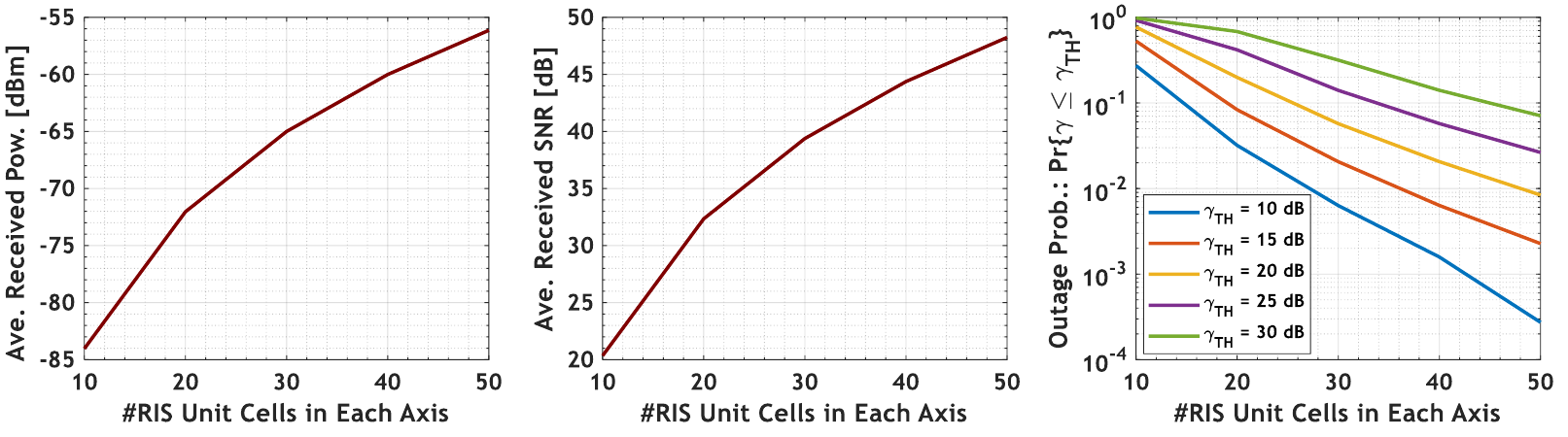}}
    \subfigure[UC-7]{ \includegraphics[width=0.98\columnwidth]{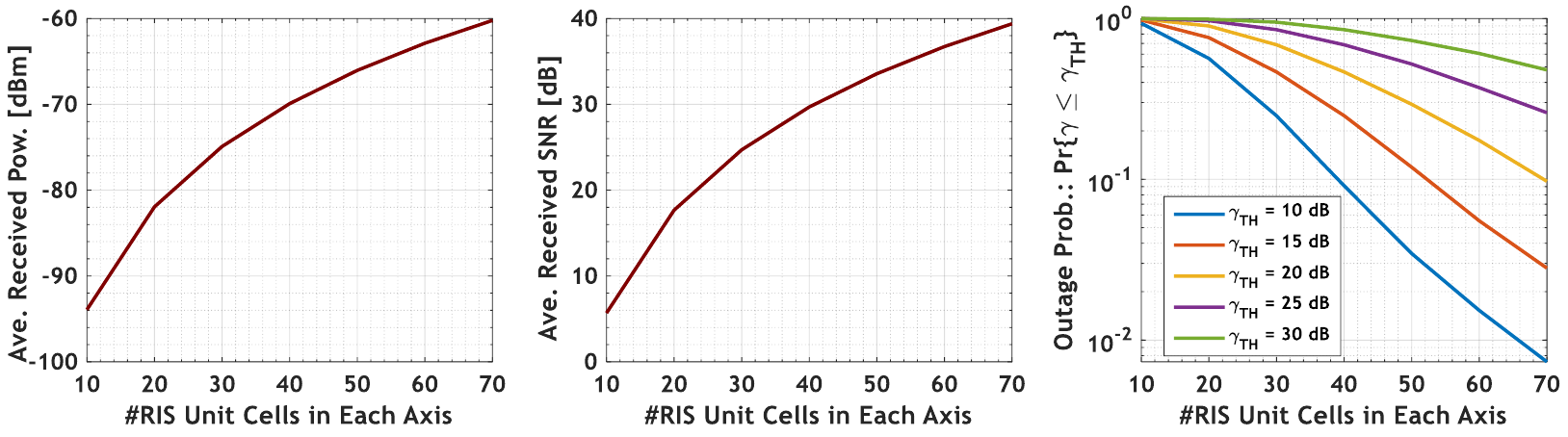}}   
    \subfigure[UC-16]{ \includegraphics[width=0.98\columnwidth]{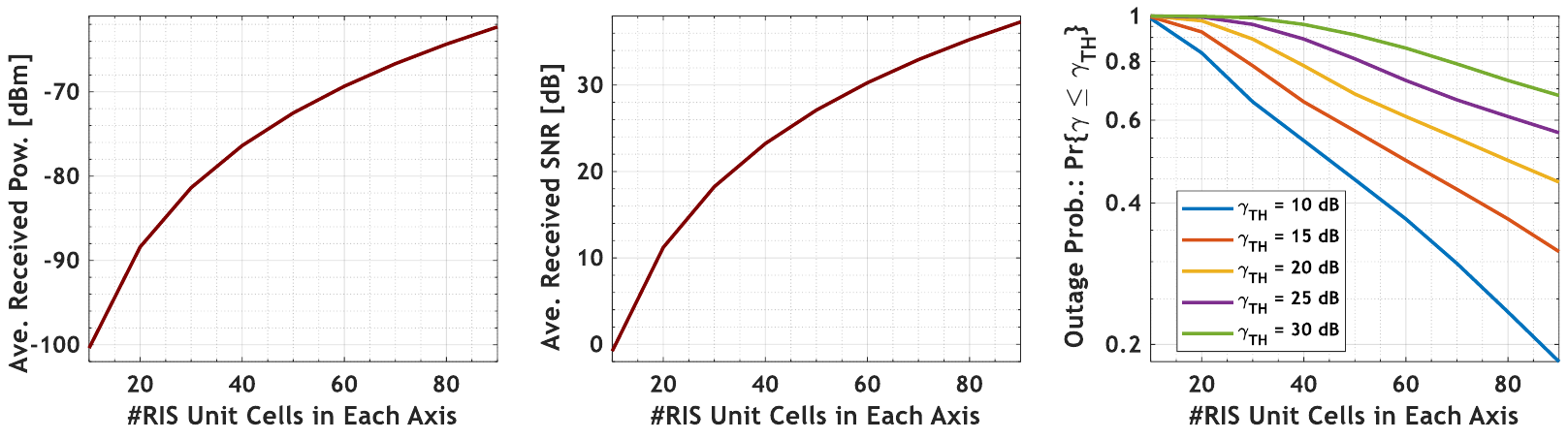}}    
    \caption{Average RIS-assisted signal power (left), SNR (middle) and OP (right) for selected UCs w.r.t. various RIS sizes}
    \label{fig:UCPerformances}
\end{figure}

\begin{figure}
    \centering
    {\includegraphics[width=0.99\columnwidth]{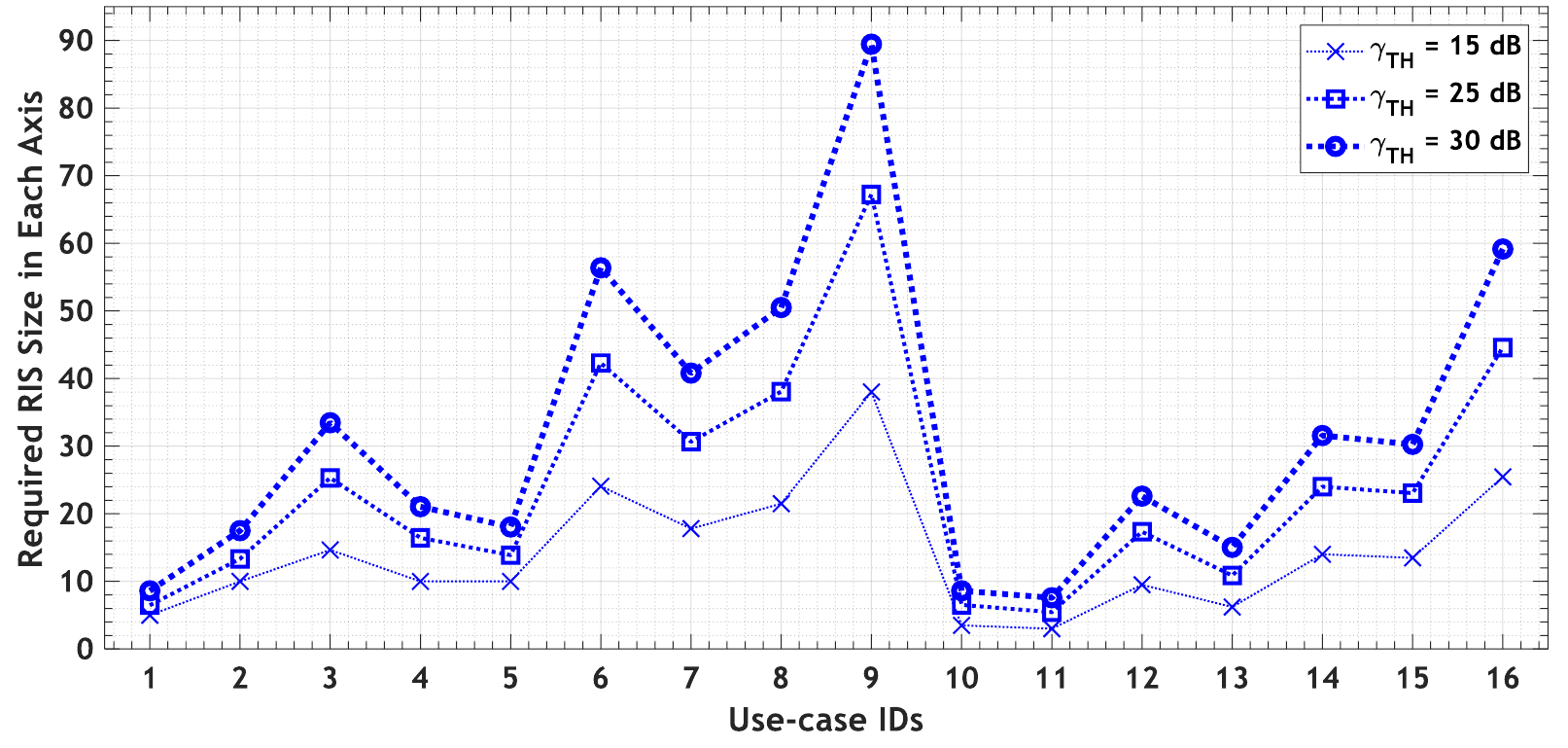}}
    \caption{Minimum RIS sizes required to achieve different SNR thresholds for all use cases}
    \label{fig:ReqRISSizes}
\end{figure}

This section presents the numerical results obtained through the simulation methodology detailed in Section II, applied to three representative use cases: UC-5, UC-7, 
and UC-16. These use cases span different deployment types, frequency bands, and geometries, allowing us to extract general trends and scenario-specific insights into the impact of RIS size on system performance. Simulation results are presented in Figs. \ref{fig:ProbabilityDensities} and \ref{fig:UCPerformances}, where each figure pair corresponds to a specific use case. For each, we show the PDF 
of average scattered power and SNR, along with the impact of varying RIS dimensions on OP. This showcases KPI variations relative to RIS dimensions, which serve as a basis for deriving general recommendations. The final summary in Fig. \ref{fig:ReqRISSizes} aggregates results from all use cases and highlights the minimum RIS sizes required to achieve various SNR thresholds, providing a broader overview of RIS size requirements and intuitive guidelines for deployment planning across frequencies and environments. 


As seen from the results given in Fig. \ref{fig:ProbabilityDensities} and Fig. \ref{fig:UCPerformances}, increasing the RIS size leads to improvements in both average power and SNR. In addition as the RIS sizes increase, the outage probabilities decrease for all usecases. Overall, the OP decreases sharply for RIS sizes beyond $20\times20$ elements, indicating a practical saturation point where additional RIS elements yield diminishing returns. UC-7 shows broader SNR and power distributions, reflective of the more complex propagation and greater angular diversity at higher altitudes and longer distances. The RIS size has a more dramatic impact on the system performance compared to UC-5, with significant improvements seen even beyond $30\times30$ arrays. The higher frequency and longer RIS-UE distances make RIS gain more sensitive to surface aperture size. Importantly, the OP remains high for small RIS dimensions, emphasizing the need for larger surfaces in such scenarios to meet service reliability requirements. In UC-16, due to the long-range BS-RIS and RIS-UE distances and sparsity of surrounding scatterers, the received power and SNR distributions show broader spreads and lower mean values, particularly for smaller RISs. Large-scale RIS deployments more than $(40\times40)$ are essential to achieve reasonable communication quality. This use case highlights the challenge of rural deployments, where the placement and scale of RISs need to be much more aggressive to compensate for geometric path loss and limited multi-path.

Fig. \ref{fig:ReqRISSizes} summarizes the minimum RIS dimensions required to meet various SNR thresholds (e.g., $5$ dB, $10$ dB, $20$ dB, $30$ dB) across all use cases. It is evident that use cases with higher operating frequencies and more challenging propagation conditions (such as UC-7 and UC-16) demand significantly larger RIS surfaces to reach the same performance targets as lower-frequency or indoor use cases (e.g., UC-5). This figure effectively captures the diversity in RIS deployment requirements and serves as a practical guideline for RIS-assisted network planning.

From the presented numerical evaluations, several key conclusions emerge. First, RIS performance improvements scale with surface size, but the marginal benefit diminishes beyond certain thresholds that are highly scenario-dependent. Second, deployment environment and frequency band play critical roles in determining optimal RIS size. Third, indoor environments, due to confined and controllable propagation, require relatively modest RIS sizes, while rural and FR3/macroscale scenarios necessitate larger surfaces to counteract long-distance losses. Lastly, the aggregate trends in the figures. provide valuable sizing references for practical RIS deployments, enabling engineers and operators to make data-driven decisions when planning RIS-supported networks across diverse 5G and 6G environments.

\section{Conclusions and Future Works}

This paper presented a practical and scenario-aware methodology for determining optimal RIS dimensions in diverse deployment environments. By introducing a more realistic scattering-based signal model and simulating various use cases, we provided insights into how RIS size impacts key performance metrics such as received power, SNR, and OP. The results demonstrate that there is no one-size-fits-all RIS design; rather, RIS dimensions must be carefully tailored to the operating frequency, geometry, and environmental characteristics of each deployment. Our findings serve as a useful guide for network planners and technology providers aiming to integrate RIS into future 6G infrastructures with cost-efficiency and performance in mind.

Future work will involve extending this framework by benchmarking RIS-assisted systems against alternative technologies (e.g., relays, repeaters), further analyzing use case-specific requirements across a broader range of frequency bands and propagation conditions, and incorporating more dynamic modeling features. While this study used a generic Friis model with a fixed path loss exponent (PLE), future enhancements will explore environment-specific PLE values, informed by measurement data, to refine simulation fidelity. Incorporating adaptive beamforming, diverse RIS architectures, and real-world validation will also be critical for evolving this simulator into a comprehensive tool to support standardization and deployment decisions.

\section*{Acknowledgment}
This study has been supported by the 1515 Frontier Research and Development Laboratories Support Program of T{\"U}B{\.I}TAK under Project 5229901 - 6GEN. Lab: 6G and Artificial Intelligence Laboratory. 
The authors express their gratitude to ZTE Corporation for sharing the KPI results from outdoor tests, which provided a valuable comparison for validating the simulation results related to RIS technology. 


\end{document}